\documentclass[aps,prx,reprint,groupedaddress,superscriptaddress,floatfix]{revtex4-2}

\usepackage[T1]{fontenc}
\usepackage{newtxtext}
\usepackage{newtxmath}

\usepackage{amsmath}
\usepackage{mathtools}
\usepackage{amsfonts}

\usepackage{color}
\usepackage{xcolor}
\usepackage[colorlinks, citecolor=blue, linkcolor=blue, urlcolor=blue]{hyperref}

\usepackage{graphicx}
\usepackage[all]{hypcap}

\usepackage{qcircuit}

\begin{document}

\title{Navigating the noise-depth tradeoff in adiabatic quantum circuits}

\author{Daniel Azses}
\affiliation{School of Physics and Astronomy, Tel Aviv University, Tel Aviv 6997801, Israel}

\author{Maxime Dupont}
\affiliation{Department of Physics, University of California, Berkeley, California 94720, USA}
\affiliation{Materials Sciences Division, Lawrence Berkeley National Laboratory, Berkeley, California 94720, USA}
\affiliation{Rigetti Computing, 775 Heinz Avenue, Berkeley, California 94710, USA}

\author{Bram Evert}
\affiliation{Rigetti Computing, 775 Heinz Avenue, Berkeley, California 94710, USA}

\author{Matthew J. Reagor}
\affiliation{Rigetti Computing, 775 Heinz Avenue, Berkeley, California 94710, USA}

\author{Emanuele G. Dalla Torre}
\affiliation{Department of Physics, Bar-Ilan University, Ramat Gan 5290002, Israel}
\affiliation{Center for Quantum Entanglement Science and Technology, Bar-Ilan University, Ramat Gan 5290002, Israel}

\begin{abstract}
    Adiabatic quantum algorithms solve computational problems by slowly evolving a trivial state to the desired solution. On an ideal quantum computer, the solution quality improves monotonically with increasing circuit depth. By contrast, increasing the depth in current noisy computers introduces more noise and eventually deteriorates any computational advantage. What is the optimal circuit depth that provides the best solution? Here, we address this question by investigating an adiabatic circuit that interpolates between the paramagnetic and ferromagnetic ground states of the one-dimensional quantum Ising model. We characterize the quality of the final output by the density of defects $d$, as a function of the circuit depth $N$ and noise strength $\sigma$. We find that $d$ is well-described by the simple form $d_\mathrm{ideal}+d_\mathrm{noise}$, where the ideal case $d_\mathrm{ideal}\sim N^{-1/2}$ is controlled by the Kibble-Zurek mechanism, and the noise contribution scales as $d_\mathrm{noise}\sim N\sigma^2$. It follows that the optimal number of steps minimizing the number of defects goes as $\sim\sigma^{-4/3}$. We implement this algorithm on a noisy superconducting quantum processor and find that the dependence of the density of defects on the circuit depth follows the predicted non-monotonous behavior and agrees well with noisy simulations. Our work allows one to efficiently benchmark quantum devices and extract their effective noise strength $\sigma$.
\end{abstract}

\maketitle

\section{Introduction}

Quantum computing has recently achieved several key milestones, including the demonstration of quantum advantage~\cite{Preskill2012,Harrow2017,Boixo2018,Arute2019,Zhong2020,Madsen2022} and the democratization of quantum hardware through cloud services. Although more and better qubits are needed to realize the full potential of quantum computers, current devices are at a turning point for the development and testing of algorithms. The key obstacle of the current hardware is the inherent noise, which limits the size of the quantum circuits that can be executed reliably~\cite{Preskill2018}. Currently, this number involves, at most, a few tens of qubits~\cite{pelofske2022quantum,niroula2022constrained}. However, it is expected to grow significantly, hopefully reaching the hundreds in the next few years.

Important examples of quantum algorithms that can be run on intermediate-scale quantum computers include adiabatic state preparation protocols~\cite{RevModPhys.90.015002} and variational algorithms~\cite{Cerezo2021,wang2021noise} such as the quantum approximate optimization algorithm (QAOA)~\cite{Farhi2014,Farhi2014b,Farhi2016}. These algorithms converge asymptotically to the correct solution with increasing circuit depth, i.e., the output becomes better with each additional layer.  However, the presence of noise leads to a fundamental trade-off between computational power and accuracy: On the one hand, the user desires to perform complex calculations and run quantum circuits with a large number of layers. On the other, increasing the circuit depth leads to an increase of the noise and, eventually, deteriorates any computational advantage. A fundamental question faced by the quantum programmer is at what point does the noise introduced by an additional circuit layer overcome its algorithmic benefit. In other words, what is the optimal number of layer to obtain the best solution in the presence of noise?

We address this question by considering an adiabatic quantum protocol that transforms an equal superposition of all basis states to a generalized Greenberger-Horne-Zeilinger (GHZ) state~\cite{Greenberger2007} $(\vert 00\ldots 0\rangle + \vert 11\ldots 1\rangle)/\sqrt{2}$. We quantify the success of our algorithm by measuring the number of defects in the final state, defined such that the GHZ state corresponds to the absence of defects, $d=0$. For a one-dimensional system of $L$ qubits, we characterize the density of defects by
\begin{equation}
    d=\frac{1}{2(L-1)}\sum\nolimits_{i=1}^{L-1}\Bigl(1-\bigl\langle Z_iZ_{i+1}\bigr\rangle\Bigr)~~\in[0,1],
    \label{eq:density_defects}
\end{equation}
where $L-1$ is the number of bonds, and $Z_i$ is the Pauli operator on qubit $i$. For example, the basis state $\vert 00001111\rangle$ has one defect, i.e., a single domain wall separating two continuous sequences of the same qubit state. In contrast, the initial state of our protocol has, on average, one defect every two bonds and corresponds to $d=1/2$. According to the adiabatic theorem, the perfect GHZ state ($d=0$) is obtained only in the limit of infinite circuit depth, while a finite-depth circuit will induce defects in the final quantum state ($d>0$). Hence, by monitoring $d$ throughout the circuit, we can estimate the success rate of our algorithm.

In absence of noise, the dependence of the number of defects $d$ on the circuit depth $N$ is governed by the Kibble-Zurek (KZ) mechanism~\cite{Kibble1976,Zurek1985,DelCampo2014}, 
which has been extensively studied in the literature~\cite{PhysRevLett.95.105701,PhysRevLett.95.245701,PhysRevA.75.052321,Dziarmaga2010,PhysRevA.83.062104,PhysRevA.100.032115,PhysRevLett.109.015701,PhysRevB.86.064304,Russomanno2016,dutta2016kibble,PhysRevB.93.075134,PhysRevLett.124.090502,PhysRevLett.123.130603,Ebadi2021,Schmitt2021}, and observed experimentally in cold atoms and superconducting qubits~\cite{Gong2016,Keesling2019,PhysRevB.106.L041109}. According to the KZ scaling, $d$ is proportional to $N^{-\alpha}$, where the exponent $\alpha>0$ is set by the critical exponents of the model. Our goal is to understand how the density of defects behaves in the presence of noise. Qualitatively, we expect the density of defects induced by the noise to be an increasing function of the circuit depth $N$. The interplay between the KZ mechanism and the noise will generically lead to a non-monotonous behavior, which we aim to characterize. Inspired by Ref.~\cite{PhysRevLett.101.175701,dutta2016kibble,PhysRevLett.124.090502}, we assume the following ansatz,
\begin{equation}
    d\simeq d_\mathrm{ideal} + d_\mathrm{noise},
    \label{eq:defects_ansatz}
\end{equation}
where $d_\mathrm{ideal}$ is the density of defects in an noiseless circuit and $d_\mathrm{noise}$ is the density of defects induced by the noise. We verify this ansatz by extensive numerical simulations, aimed at determining the dependence of $d$ on the number of layers $N$ and on the noise strength $\sigma$. As a key result of our calculations, we predict the circuit depth for which the density of defects is minimal. By comparing experimental results with an empirical noise model, we propose a method to benchmark quantum computers by extracting their corresponding noise strength $\sigma$.

\section{Model, definitions, and methods}

In this work, we focus on a paradigmatic quantum model, namely the one-dimensional Ising model in a transverse field, also known as the quantum Ising model. This model interpolates linearly between the paramagnetic (PM) and ferromagnetic (FM) Hamiltonians,
\begin{align}
    H^\mathrm{PM}=-\sum\nolimits_{i=1}^{L}X_i~~~\mathrm{and}~~~H^\mathrm{FM}=-\sum\nolimits_{i=1}^{L-1}Z_iZ_{i+1},
    \label{eq:hamiltonian}
\end{align}
where $X_i$ and $Z_i$ are the Pauli operators on qubit $i$. The quantum processor is initially prepared in the ground state of $H^\mathrm{PM}$ by applying individual Hadamard gates on all qubits. We, then, implement an adiabatic protocol going from $H^\mathrm{PM}$ to $H^\mathrm{FM}$. At the end of the protocol, we measure the qubits in the computational basis (the $z$-component of the spin) and compute the density of defects according to Eq.~\eqref{eq:density_defects}.

\begin{figure}[t]
    \begin{tabular}{cc}
        \begin{tabular}{c}
            (a)\\\\
            \raisebox{3.1cm}{\Qcircuit @C=0.5em @R=1.2em {	
                & \qw & \multigate{1}{\mathtt{R_{zz}}} & \qw & \qw & \gate{\mathtt{R_x}} & \qw\\
                & \qw & \ghost{\mathtt{R_{zz}}} & \multigate{1}{\mathtt{R_{zz}}} & \qw & \gate{\mathtt{R_x}} & \qw\\
                & \qw & \multigate{1}{\mathtt{R_{zz}}} & \ghost{\mathtt{R_{zz}}}  & \qw & \gate{\mathtt{R_x}} & \qw\\
                & \qw & \ghost{\mathtt{R_{zz}}} & \qw & \qw & \gate{\mathtt{R_x}} & \qw
            }}
        \end{tabular}&
        \begin{tabular}{c}
            (b)\\ 
            \includegraphics[width=0.52\columnwidth]{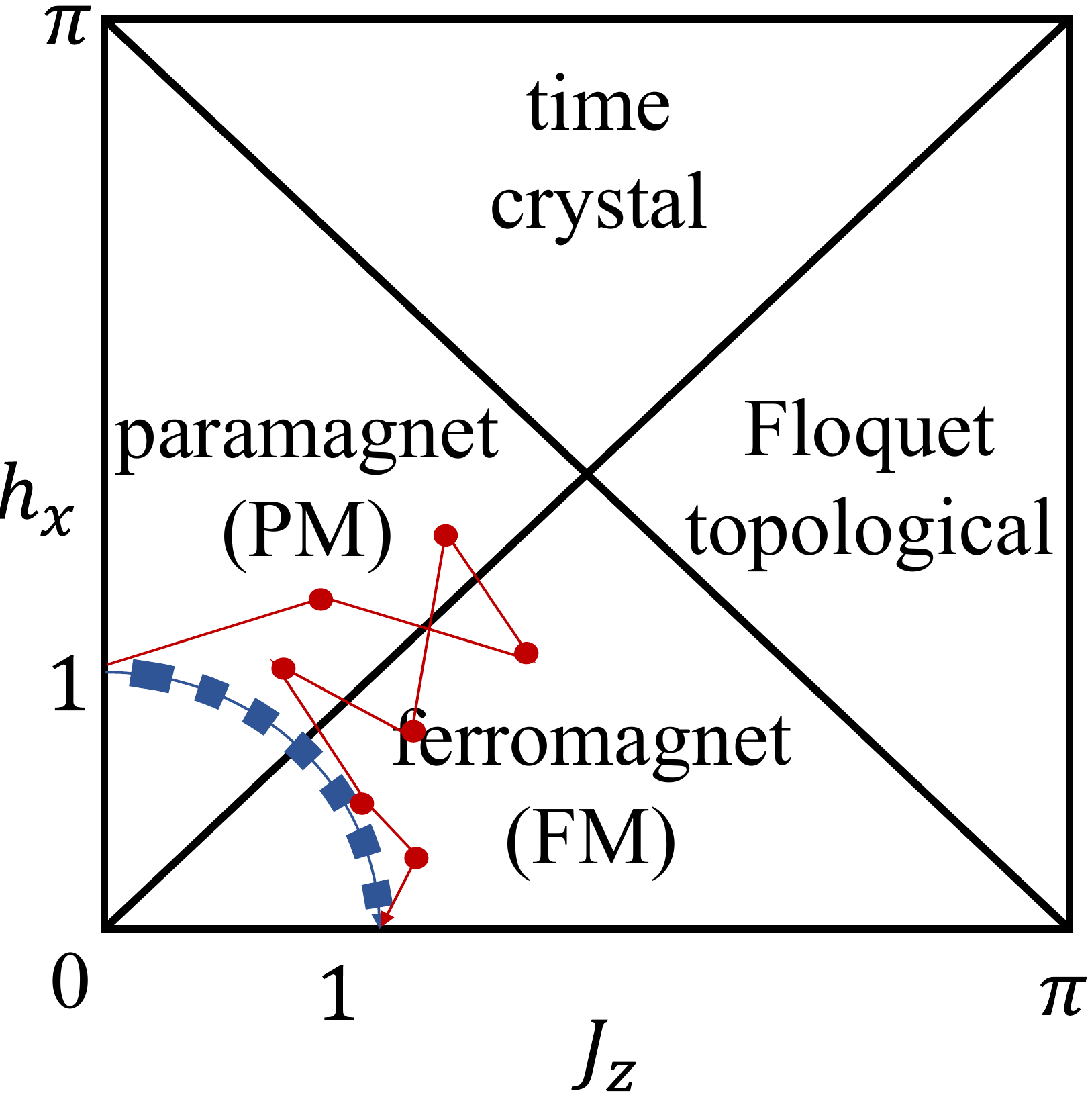}
        \end{tabular}
    \end{tabular}
    \caption{(a) Quantum circuit representing a single step of our protocol for $L=4$ qubits, consisting of parametric gates $\mathtt{R_{zz}}$ and $\mathtt{R_x}$ implementing the unitary evolution, controlled, respectively, by $H^\mathrm{FM}$ and $H^\mathrm{PM}$, see Eqs.~\eqref{eq:hamiltonian} and~\eqref{eq:unitary_step}. (b) Phase diagram of the one-dimensional Floquet quantum Ising model generated by a periodic application of Eq.~\eqref{eq:unitary_step} with fixed $h^x$ and $J^z$. The blue path corresponds to the protocol considered in this work, while the red path to a possible implementation of a QAOA algorithm with the same number of steps, $N=7$.}
    \label{fig:phasediagram}
\end{figure}

The simplest way to perform an adiabatic evolution is to drive the system according to the Hamiltonian $H(s)=(1-s)H^\mathrm{PM}+sH^\mathrm{FM}$ with  $s$ slowly increasing over time from 0 to 1. This method corresponds to adiabatic quantum computation and fits analog quantum devices~\cite{RevModPhys.90.015002}. To work with digital quantum computers, we perform a first-order Suzuki-Trotter decomposition of the adiabatic protocol, and split it into $n=1,\ldots,N$ smaller steps, each described by the unitary operator
\begin{equation}
	U_n=\exp\Bigl(-ih^x_nH^\mathrm{PM}\Bigr)\exp\Bigl(-iJ^z_nH^\mathrm{FM}\Bigr),
	\label{eq:unitary_step}
\end{equation}
with step-dependent parameters $h^x_n$ and $J^z_n$ following a chosen procedure. This protocol can be implemented as a quantum circuit, see Fig.~\ref{fig:phasediagram}(a). Thanks to a mapping to free fermions through a Jordan-Wigner transformation, the resulting time evolution can be computed in polynomial time by classical means~\cite{kitaev2001unpaired,aguado2017majorana,Terhal2002Classical,Wimmer2012Algorithm,Azses2021Observing}. If $h^x$ and $J^z$ are kept fixed, one obtains a Floquet system where $H^\mathrm{PM}$ and $H^\mathrm{FM}$ alternate in time~\cite{kick}. The phase diagram of this model was studied in Ref.~\cite{khemani2016phase}, see Fig.~\ref{fig:phasediagram}(b): In addition to the FM and PM phases, the model has a Floquet topological and a discrete time crystal phases~\cite{else2016floquet,phase_diagram}. These phases have been observed experimentally in digital quantum computers, in Refs.~\cite{Azses2021Observing},~\cite{xu2021realizing}, and~\cite{mi2021observation}, giving rise to significant public interest. 

In this work, we vary $h^x_n$ and $J^z_n$ along the rounded path depicted by the blue squares in Fig.~\ref{fig:phasediagram}(b)~\cite{Azses2021Observing}, formally described by $h^x_n=\cos\theta_n$ and $J^z_n=\sin\theta_n$, with $\theta_n=n\pi/[2(N+1)]$ and $n=1,2,\ldots N$. This protocol connects adiabatically $H^{\rm PM}$ to $H^{\rm FM}$ and crosses a quantum phase transition at the step $n=N/2$. Incidentally, we observe that the time evolution of Eq.~\eqref{eq:unitary_step} is the building block of the QAOA, where $\cos\theta_n$ and $\sin\theta_n$ are substituted by variational parameters, as schematically drawn by the red dots in Fig.~\ref{fig:phasediagram}(b).

\section{Defects in noiseless and noisy adiabatic protocols}

\subsection{Noiseless case: the Kibble-Zurek scaling}

We first turn our attention to the density of defects $d_\mathrm{ideal}$ induced by a finite number of layers $N$ in a noiseless scenario. In general, because the initial and final states belong to different phases of matter (paramagnetic versus ferromagnetic), a phase transition is expected to happen on the way~\cite{Sachdev2011}. If the transition is second order, in the vicinity of the transition point the spectral gap between the ground state and the first excited state closes as $\sim L^{-z}$, where $z$ the dynamical critical exponent. This minimal gap is the relevant energy scale for an adiabatic evolution and dictates the maximal velocity $u$ at which the protocol can be carried such that the system remains close to its instantaneous ground state~\cite{RevModPhys.90.015002}. According to the KZ mechanism, the density of defects at the end of the protocol should scale as $d\sim u^{z/(z+\nu)}$, where $\nu$ is the correlation-length critical exponent~\cite{DelCampo2014}. In the context of gate-based quantum computing, the protocol is performed using $N$ discrete steps. For fixed initial and final points, the velocity $u$ is inversely proportional to $N$ and one expects $d_\mathrm{ideal}\sim N^{-z/(z+\nu)}$. The applicability of the KZ scaling to the discrete evolution of integrable models was first demonstrated in Ref.~\cite{Russomanno2016} and dubbed Floquet-Kibble-Zurek mechanism. This result can be understood by noting that the KZ scaling is determined by the low-frequency component of the single-particle spectrum, which is identical for the continuous and discrete (Floquet) adiabatic evolution.

\begin{figure}[t]
	\includegraphics[width=\columnwidth]{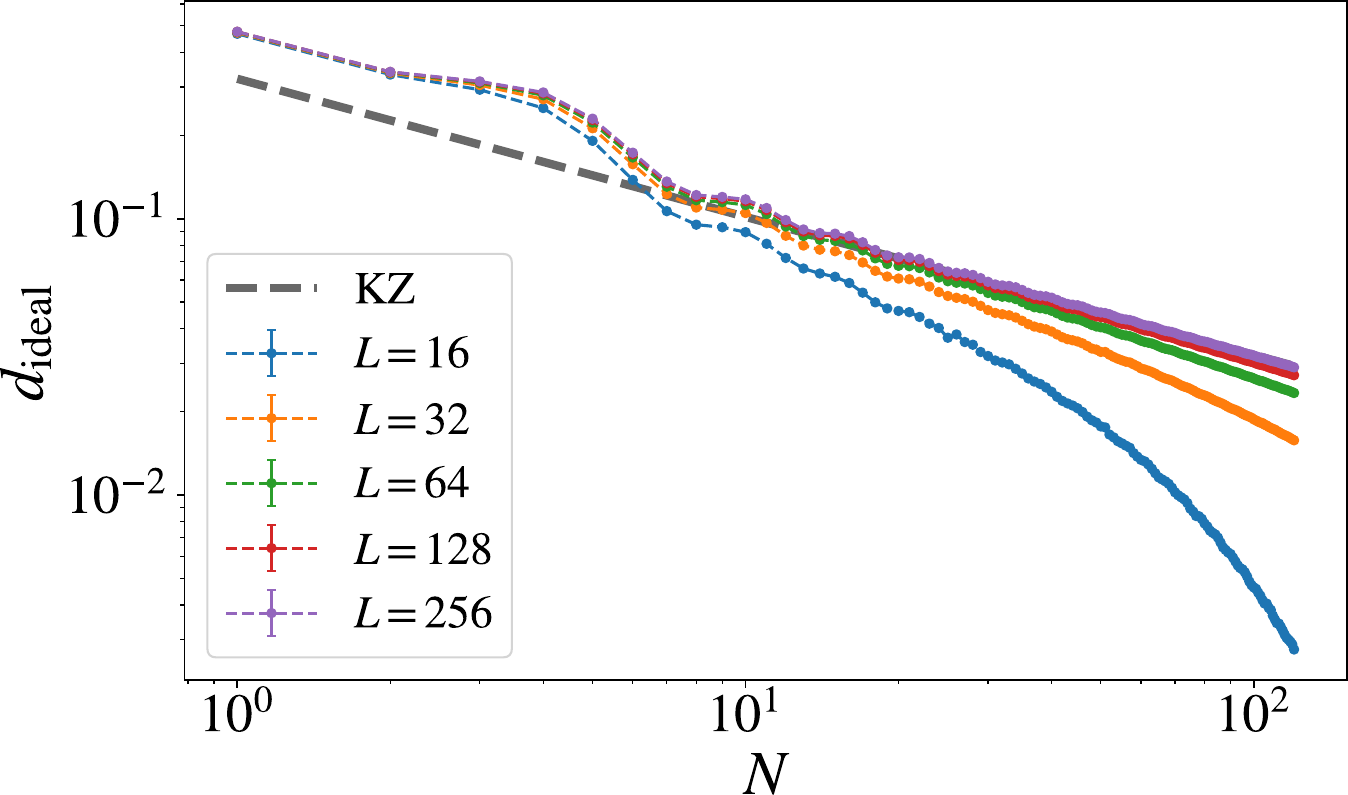}
 	\caption{Density of defects in the ideal case as a function of the number of steps $N$, for various system sizes $L$. For large system sizes, $d$ follows the KZ scaling  $d_\mathrm{ideal}=0.3213(3)/\sqrt{N}$ (dashed line, data fitted for $N>50$ on $L=256$), while for small systems, $L\ll N^2$, the suppression becomes exponential.}
	\label{fig:FIG7}
\end{figure}

For the one-dimensional quantum Ising model of Eq.~\eqref{eq:hamiltonian}, the correlation-length and dynamical exponents are $\nu=z=1$~\cite{cardy1996scaling,Sachdev2011}, leading to a square-root decay of the density of defects with the number of layers, i.e., $d_\mathrm{ideal}\sim 1/\sqrt{N}$. This scaling is verified in Fig.~\ref{fig:FIG7} for  the largest system size. For smaller system sizes, we observe a systematic deviation from the expected power-law, which can be understood as follows. The density of defects is associated with a defect-free length scale $\xi_\mathrm{ideal}\sim d_\mathrm{ideal}^{-1}\sim\sqrt{N}$. The KZ scaling is valid in the limit where the system size $L$ is much larger than $\xi_\mathrm{ideal}$, or equivalently $N\ll L^2$, such that finite-size effects can be neglected. If the number of steps exceeds this limit, one expects to recover the result of a finite system, where the density of excitations (defects) decays exponentially with $N$.

\subsection{Noisy case: Step-dependent noise and static disorder}
\label{sec:noisy_model}

The main goal of this work is to study the KZ mechanism in a noisy environment. We introduce noise in the form of the following modifications to Eq.~\eqref{eq:hamiltonian},
\begin{equation}
    H^\mathrm{PM}_n=-\sum\nolimits_{i=1}^{L}\Bigl[1+\eta_{n,i}\bigl(\sigma_\mathrm{noise}\bigr)+\eta_i\bigl(\sigma_\mathrm{disorder}\bigr)\Bigr]X_i,
    \label{eq:hamiltonian_noise_pm}
\end{equation}
and,
\begin{equation}
    H^\mathrm{FM}_n=-\sum\nolimits_{i=1}^{L-1}\Bigl[1+\eta_{n,i}\bigl(\sigma_\mathrm{noise}\bigr)+\eta_i\bigl(\sigma_\mathrm{disorder}\bigr)\Bigr]Z_iZ_{i+1},
    \label{eq:hamiltonian_noise_fm}
\end{equation}
where the above Hamiltonians now have an explicit dependence on the Floquet evolution step $n$. Here, $\eta(\sigma)$ is a random variable, normally distributed with mean zero and standard deviation $\sigma$, characterizing the strength of the noise. These equations introduce two types of randomness: The first one, referred to as noise, is qubit- and step-dependent. The second, called disorder, is only qubit-dependent and does not vary at each step. 
Equations~\eqref{eq:hamiltonian_noise_pm} and~\eqref{eq:hamiltonian_noise_fm} are analogous to the models introduced in Refs.~\cite{dutta2016kibble,PhysRevLett.124.090502,PhysRevResearch.2.033369,King2022} in the context of continuous-time adiabatic annealing. Specifically, Ref.~\cite{dutta2016kibble} considered a model of spatially uniform noise, described by a time-dependent field that couples to $J_z\to J_z(t)$, rather than to the individual qubits. Our noise model is analogous to the infinite temperature limit of the Ohmic bath considered in Ref.~\cite{PhysRevResearch.2.033369}, and our disorder to the random magnetic field used in Ref.~\cite{King2022} to account for kink correlations. We will comment more on these analogies below.

To describe phenomenologically the effects of noise and disorder, we  extend the ansatz of Eq.~\eqref{eq:defects_ansatz} regarding the different contributions to the density of defects to,
\begin{equation}
    d_\mathrm{noise}~\longrightarrow~d_\mathrm{noise} + d_\mathrm{disorder},
    \label{eq:defects_noise_plus_disorder}
\end{equation}
with $d_\mathrm{noise}$ the contribution of the step-dependent noise specifically and $d_\mathrm{disorder}$ the contribution of the static disorder. We seek to find a functional form for the two contributions as a function of the randomness strength and number of steps. The modifications of Eqs.~\eqref{eq:hamiltonian_noise_pm} and~\eqref{eq:hamiltonian_noise_fm} conserve the free fermionic nature of the circuit, which allows us to use the same technique as the noiseless case for investigating the system.

\subsubsection{Step-dependent noise}

\begin{figure}[t]
	\includegraphics[width=1\columnwidth]{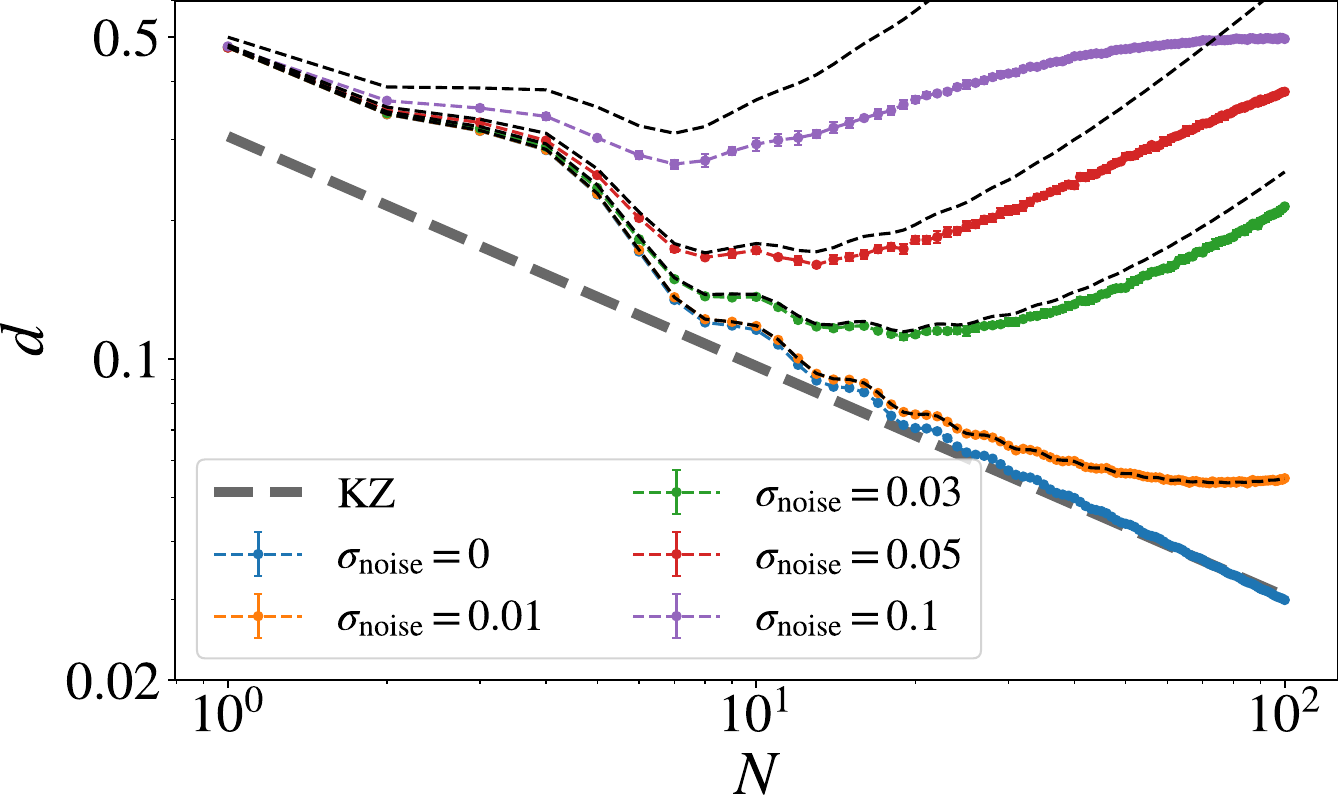}
	\caption{Density of defects $d$ for $L=120$ for various noise strengths. The purple dashed line represents the KZ scaling $d\equiv d_\mathrm{ideal}=0.3046(5)/\sqrt{N}$ for the ideal case (data fitted for $N>50$). The black dashed curves are obtained using Eq.~\eqref{eq:density_defects} where $d_\mathrm{ideal}$ is given by the noiseless data and $d_\mathrm{noise}=2.499(4)\sigma^2N$ (data fitted for $N>10$ and $\sigma_\mathrm{noise}=10^{-2}$). Each data point is generated from the average over ten random samples.}
	\label{fig:FKZnoisy_L50}
\end{figure}

We first consider the step-dependent noise case, setting $\sigma_\mathrm{noise}>0$ and $\sigma_\mathrm{disorder}=0$. In the limit of an infinite number of steps, for any noise strength, the circuit falls into the class of random unitary free fermion circuits~\cite{Dias2021} and leads to a random Gaussian state. It follows that the total density of defects according to Eq.~\eqref{eq:density_defects} for such a state is $d=1/2$. Here, we are interested in the regime before the defects density saturates. Figure~\ref{fig:FKZnoisy_L50} shows the density of defects as a function of the number of steps for various values of the noise strength $\sigma_\mathrm{noise}\in[0, 0.1]$. At small but nonzero noise, the data initially follows the ideal case ($\sigma_\mathrm{noise}=0$) and decreases in the first steps as $\sim 1/\sqrt{N}$, but then deviates and begins to increase. The deviation from the ideal case happens for a smaller and smaller number of steps as the noise strength increases. We observe a noise-dependent minimum as a function of the number of steps, corresponding to the optimal number of steps $N_\mathrm{opt}$ that minimizes the number of defects in the noisy adiabatic circuit. We come back to this minimum in Sec.~\ref{sec:optimal_depth}.

\begin{figure}[t]
	\includegraphics[width=0.9\linewidth]{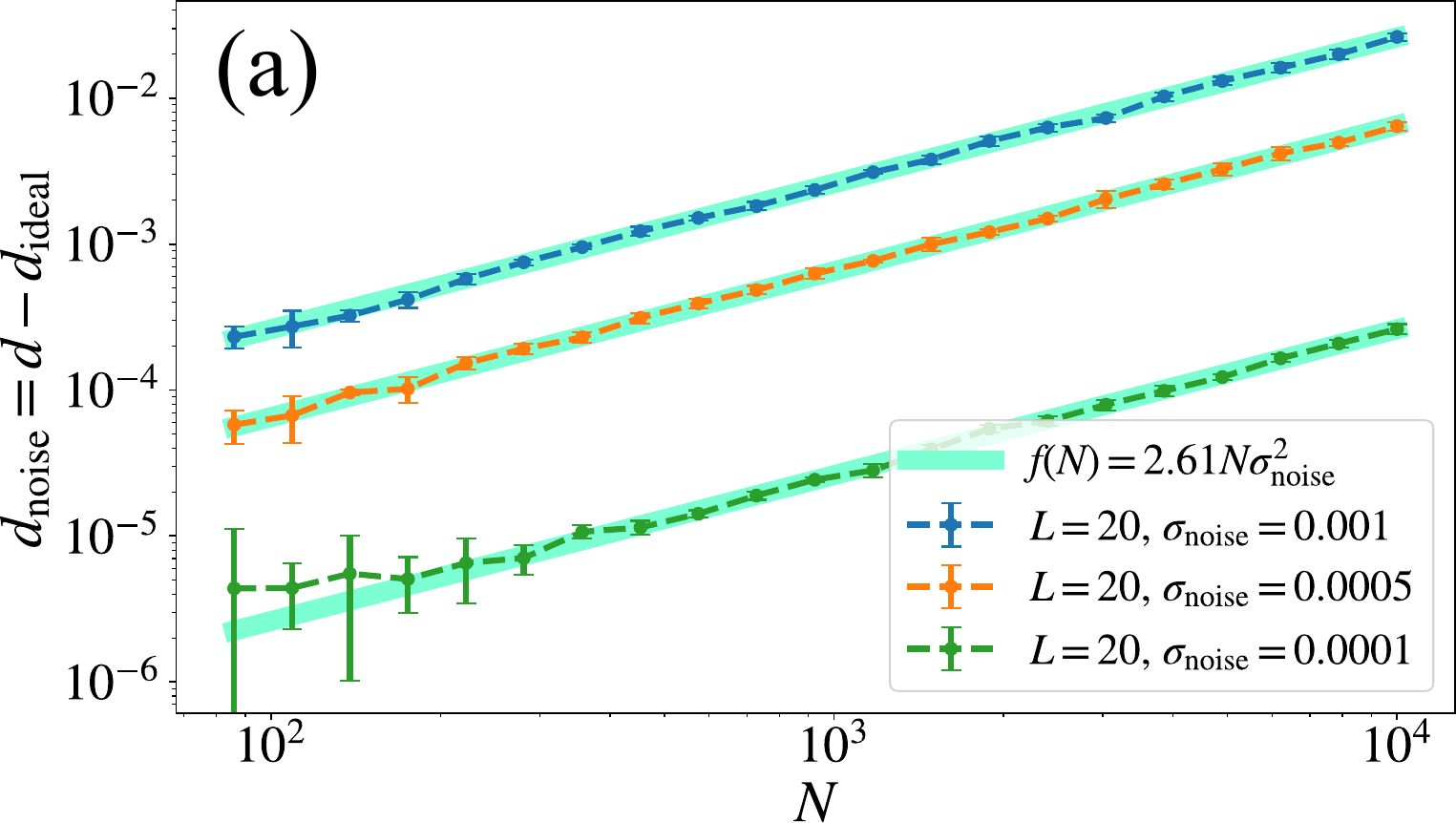}\\
	\includegraphics[width=0.9\linewidth]{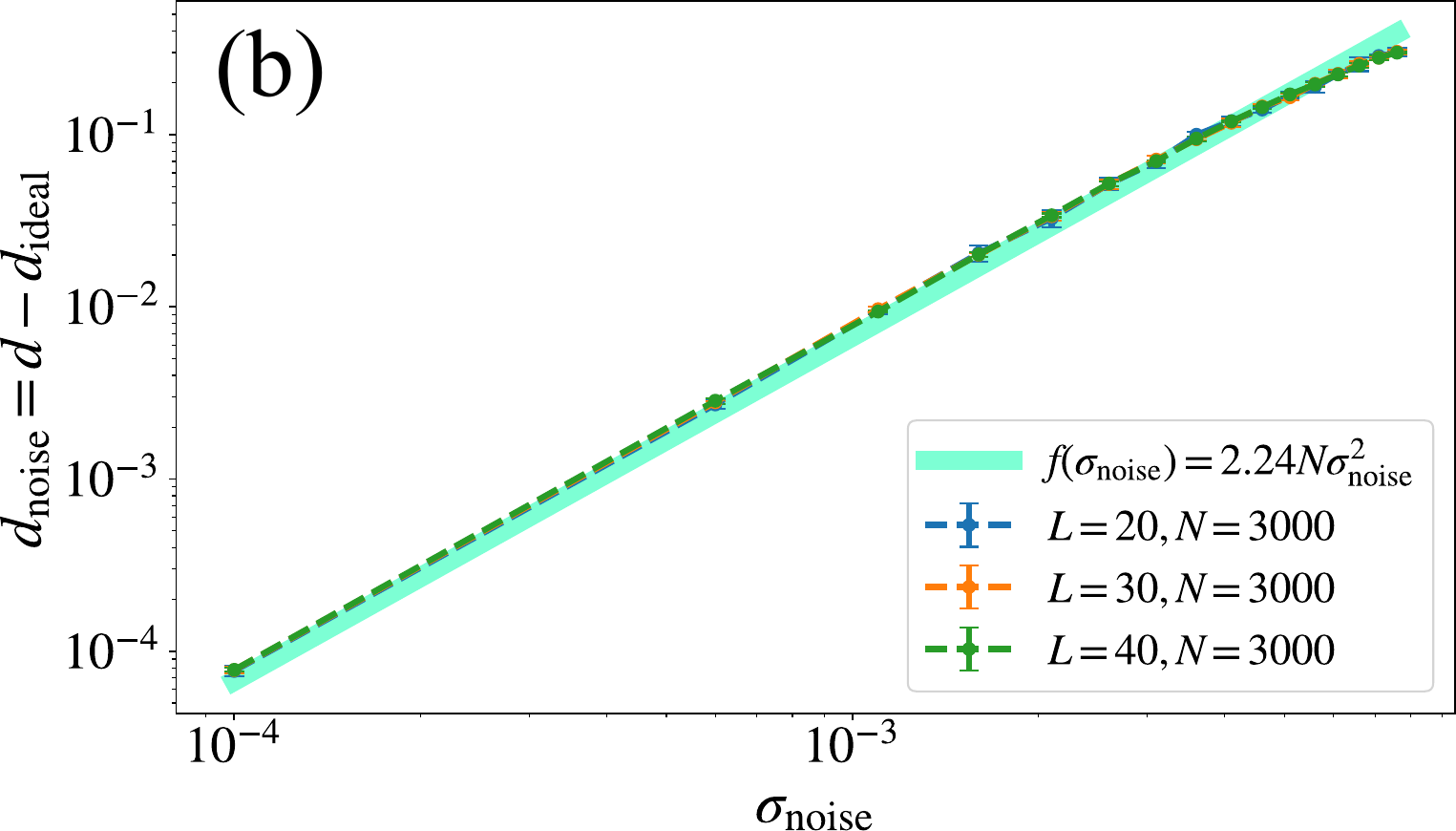}\\
	\includegraphics[width=0.9\linewidth]{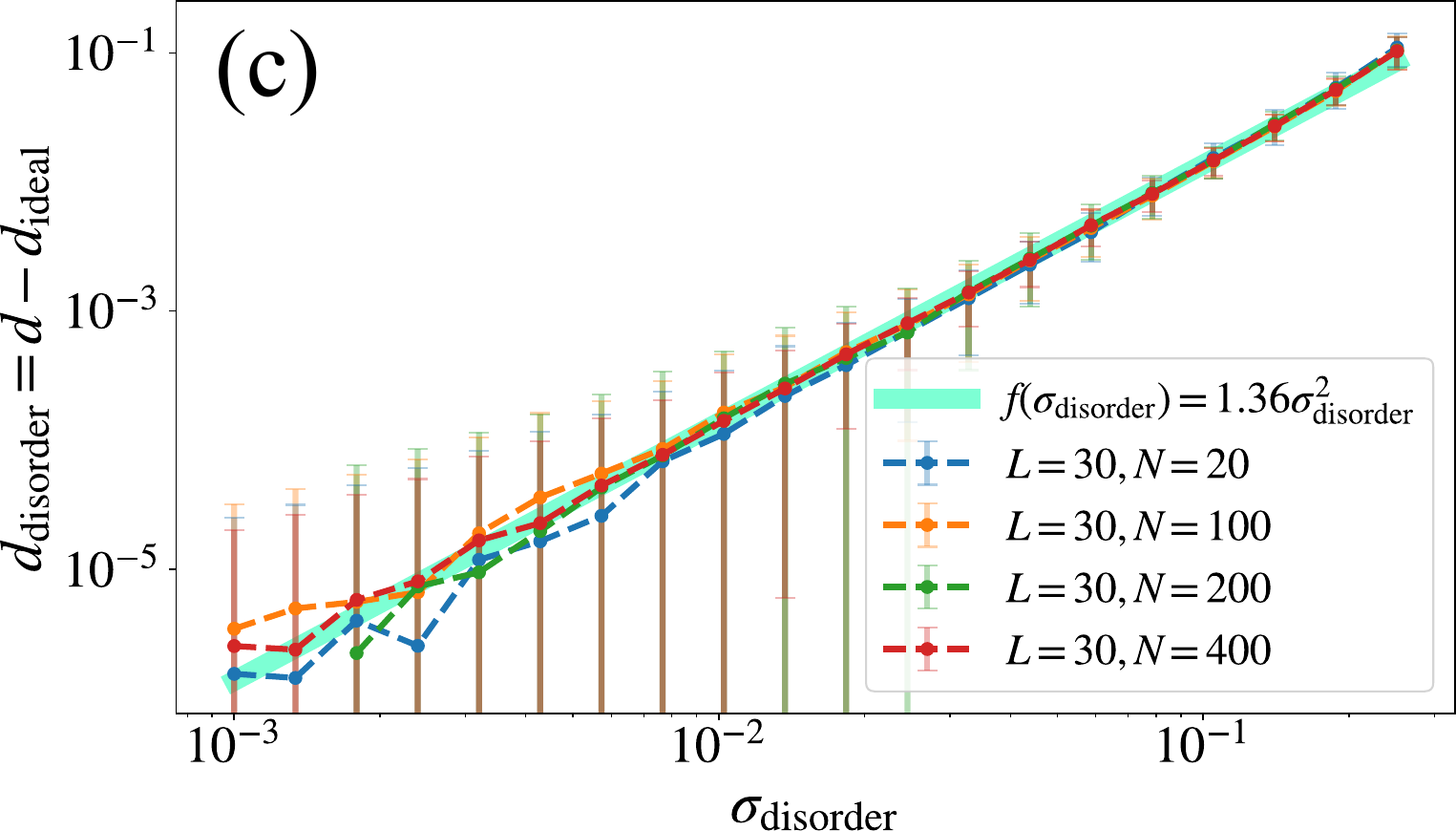}
	\caption{(a) Density of defects in the presence of noise, as a function of the number of steps $N$, for various noise strengths $\sigma_\mathrm{noise}$. We consider a system size of $L=20$ and average the data over five random realizations. We observe a linear dependence with $N$. (b) Density of defects, as a function of the noise strength, for a fixed number of steps $N=3\times 10^3$. There is no size dependence as the data for $L=20$, $L=30$, and $L=40$ fall onto each others. The data is averaged over five random realizations. We find an algebraic increase with the noise strength $\sim\sigma^2_\mathrm{noise}$. (c) Density of defects in presence of static disorder, as a function of the disorder strength $\sigma_\mathrm{disorder}$, for various number of steps $N$ (data averaged over five hundred random realizations). We find no dependence with $N$ and checked that there is no system size dependence either (data not shown). We observe that the density of disorder-induced defects grows as $\sim\sigma_\mathrm{disorder}^2$.}
	\label{fig:FKZnoise_scaling}
\end{figure}

If the ansatz of Eq.~\eqref{eq:defects_ansatz} is valid, we can isolate the noise-induced defects by subtracting the adiabatic contribution coming from the use of a finite number of steps $N$, $d_\mathrm{noise}\simeq d-d_\mathrm{ideal}$. For $L=20$ at fixed value of the noise strength $\sigma_\mathrm{noise}=10^{-3}$, we compute the dependence of $d_\mathrm{noise}$ with the number of steps $N$ and plot it in Fig.~\ref{fig:FKZnoise_scaling}(a). As expected, the noise increases the density of defects. We find that it grows linearly as $d_\mathrm{noise}\sim N$, to be compared with $d_\mathrm{ideal}\sim 1/\sqrt{N}$ in the ideal noiseless case. We repeat the procedure for a fixed number of steps $N=3\times 10^3$ ($L=40$) and vary the noise strength $\sigma_\mathrm{noise}$. The data is displayed in Fig.~\ref{fig:FKZnoise_scaling}(b) and shows an algebraic dependence compatible with $d_\mathrm{noise}\sim\sigma_\mathrm{noise}^2$. Put together, the two results lead to,
\begin{equation}
    d_\mathrm{noise}\simeq a_\mathrm{noise}N\sigma_\mathrm{noise}^2,
    \label{eq:defects_noise}
\end{equation}
with $a_\mathrm{noise}=2.61(2)$ and $a_\mathrm{noise}=2.24(8)$ evaluated by least-square fitting from the data of Figs.~\ref{fig:FKZnoise_scaling}(a) and~\ref{fig:FKZnoise_scaling}(b), respectively. We attribute the negligible difference between these two values to finite-size effects and consider the average  $a_\mathrm{noise}=2.42(8)$ in the following. Equation~\eqref{eq:defects_noise} is analogous to the results obtained in Refs.~\cite{PhysRevB.86.060408} and~\cite{PhysRevB.89.024303} for the case of a sudden quench, according to the identification of $\xi_\mathrm{noise}=d^{-1}_\mathrm{noise}\sim(N\sigma_\mathrm{noise}^2)^{-1}$. In Fig.~\ref{fig:FKZnoisy_L50} we compare the numerical data with $d=d_\mathrm{ideal}+d_\mathrm{noise}$, where $d_\mathrm{noise}$ is given by Eq.~\eqref{eq:defects_noise}. A good agreement is found with the simulations, including for the position of the minimum of defects as a function of the number of steps. Note that the expression of Eq.~\eqref{eq:defects_noise} does not account for the saturation of the density of defect to $d\lesssim 1/2$ as $N\to+\infty$, which is, therefore, not captured.

We now present an intuitive argument aimed at explaining the scaling of Eq.~\eqref{eq:defects_noise}. For simplicity, we consider a single qubit in the $\vert{0}\rangle$ state and to which one applies $N$ random fields of average intensity $\sigma_\mathrm{noise}$ in the $x$ direction: $\prod_{n=1}^N\exp[-i\eta_n(\sigma_\mathrm{noise})X]\vert{0}\rangle$. The resulting dynamics corresponds to a random walk in the intersection of the Bloch sphere with the YZ plane. If we denote by $\theta$ the angle of the qubit with respect to its initial state, we find that at step $N$, the variance of $\theta$ will be given by $\mathbb{E}[\theta^2]=\sigma_\mathrm{noise}^2N$. Thus, the probability to find the qubit in the $\vert{1}\rangle$ state is $d_\mathrm{noise}=1-\mathbb{E}[\cos\theta]=1-\exp(-N\sigma_\mathrm{noise}^2/2)$, where we used the central limit theorem along with properties of Gaussian variables. For small $\sigma_{\rm noise}$, one obtains $d_\mathrm{noise}=N\sigma_\mathrm{noise}^2/2$, in agreement with the result of Eq.~\eqref{eq:defects_noise}.

\subsubsection{Static disorder}

We now turn our attention to the static disorder case with $\sigma_\mathrm{disorder}>0$ and $\sigma_\mathrm{noise}=0$. The presence of disorder and the free-fermionic nature of the model leads to the Anderson localization of all single-particle states~\cite{PhysRev.109.1492,RevModPhys.80.1355}, and thus all eigenstates. In one dimension, the localization induces a length scale $\xi_\mathrm{loc}\sim\sigma_\mathrm{disorder}^{-2}$~\cite{Giamarchi_1987,PhysRevB.37.325}. The ground state of the model is still characterized by a phase transition between a paramagnetic and ferromagnetic phases. However, instead of belonging to the Ising universality class in $1+1$ dimensions~\cite{cardy1996scaling,Sachdev2011}, it proceeds via an infinite-randomness critical point~\cite{PhysRevB.51.6411}. This transition differs from the disorder free case for several reasons: (i) At the critical point, all excited eigenstates are localized, with only the ground state showing a diverging length scale; (ii) the dynamical exponent takes the value of $z=\infty$, meaning that the relationship between energy and length scales at the transition is exponential instead of being a conventional power-law; (iii) close to such a transition, physical observables $O$ can behave differently depending if one looks at the disorder average value $\mathbb{E}[O]$ or the typical value $\ln(\mathbb{E}[\exp O])$. To the best of our knowledge, there are no analytical or numerical studies of the KZ mechanism across this type of transition.

In the present study, we restrict ourselves to the limit of small $\sigma_\mathrm{disorder}$. In this limit, the disorder leads to perturbative corrections that are unaffected by the asymptotic properties of the critical point. We find that one can effectively isolate the defects induced by the disorder through $d_\mathrm{disorder}\simeq d-d_\mathrm{ideal}$, with $d_\mathrm{ideal}$ the density of defects in the ideal disorder-free case. In Fig.~\ref{fig:FKZnoise_scaling}(c), we show that the density of disorder-induced defects is independent of the number of steps $N$ and only depends on the disorder strength,
\begin{equation}
    d_\mathrm{disorder}\simeq a_\mathrm{disorder}\sigma_\mathrm{disorder}^2,
    \label{eq:defects_disorder}
\end{equation}
with $a_\mathrm{disorder}=1.36(5)$ a fitting parameter evaluated by least-square fitting on the $N=20$ data of Fig.~\ref{fig:FKZnoise_scaling}(c). The scaling of Eq.~\eqref{eq:defects_disorder} can be explained by the scaling of the localization length $d_\mathrm{disorder}\sim\xi_\mathrm{loc}^{-1}\sim\sigma_\mathrm{disorder}^{2}$.

\subsection{Minimizing defects: Optimal circuit depth}
\label{sec:optimal_depth}

\begin{figure}[t]
	\includegraphics[width=\columnwidth]{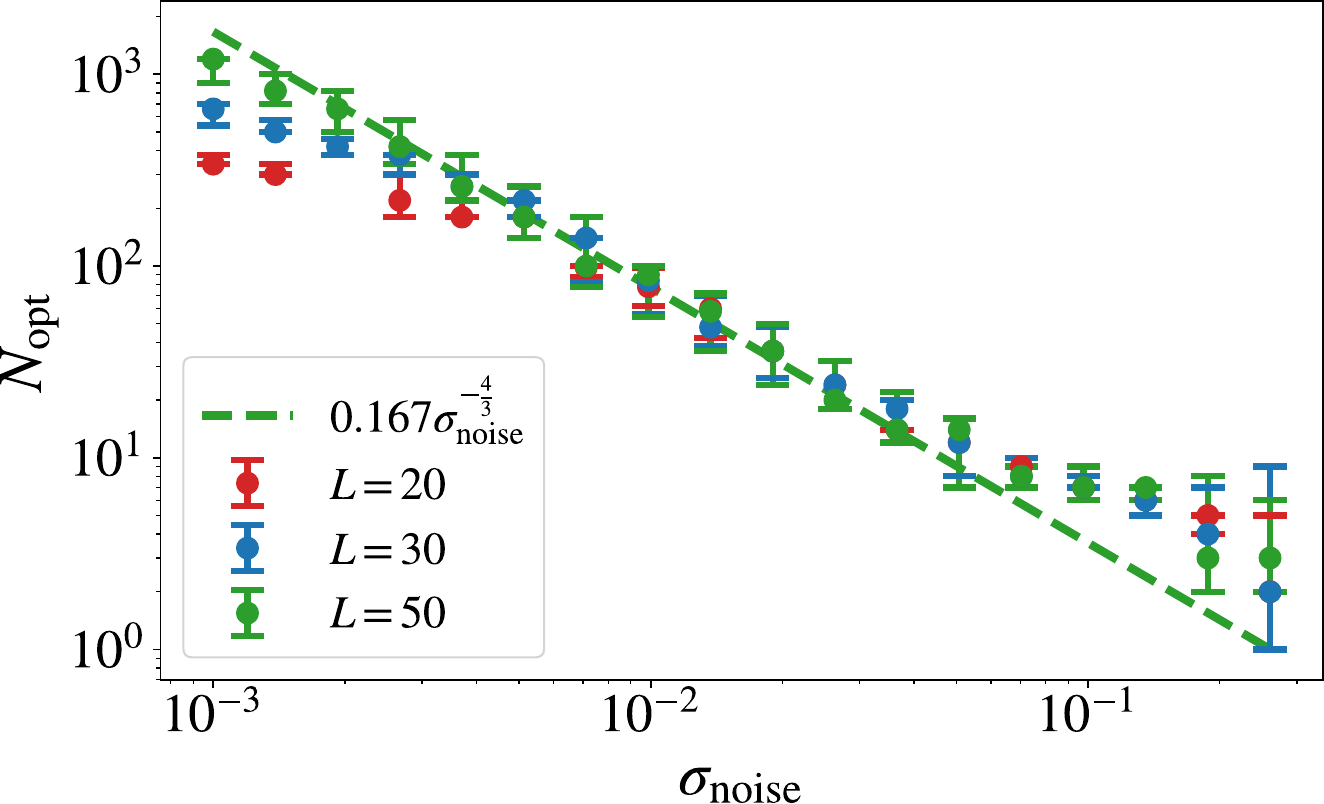}
	\caption{Optimal number of steps, as a function of the noise strength. The error bars correspond to the optimal value plus/minus $5\%$. Each data point $N_\mathrm{opt}$ has been computed from curves averaged over $10$ random realizations. We find that the optimal number of steps follows Eq.~\eqref{eq:opt_Nsteps} with $N_\mathrm{opt}=0.137(9)\sigma_\mathrm{noise}^{-4/3}$ for $L=20$, $N_\mathrm{opt}=0.164(9)\sigma_\mathrm{noise}^{-4/3}$ for $L=30$, and $N_\mathrm{opt}=0.167(7)\sigma_\mathrm{noise}^{-4/3}$ for $L=50$.}
	\label{fig:FKZoptimal_L50}
\end{figure}

In the previous section, we established the validity of the ansatz of Eq.~\eqref{eq:defects_ansatz} concerning the density of defects: The total density of defects is well-described by the sum of individual sources of defects,
\begin{equation}
    d\simeq \frac{a_\mathrm{ideal}}{\sqrt{N}} + a_\mathrm{noise}N\sigma_\mathrm{noise}^2 + a_\mathrm{disorder}\sigma_\mathrm{disorder}^2.
    \label{eq:defects_total}
\end{equation}
with the first term corresponding to the ideal scenario described by the Kibble-Zurek mechanism and the two following terms describing the noise- and disordered-induced defects, respectively. The functional form of Eq.~\eqref{eq:defects_total} versus the number of steps makes the determination of the optimal number of steps $N_\mathrm{opt}$ minimizing the total density of defects straightforward,
\begin{equation}
    \left.\frac{\partial d}{\partial N}\right|_{N=N_{\rm opt}} =0~~\Rightarrow~~N_\mathrm{opt}\simeq\sigma_\mathrm{noise}^{-4/3}\left(\frac{a_\mathrm{ideal}}{2a_\mathrm{noise}}\right)^{2/3}.
    \label{eq:opt_Nsteps}
\end{equation}
This result follows the same scaling law predicted in Refs.~\cite{dutta2016kibble,PhysRevLett.124.090502} for the case of a spatially uniform noise. Remarkably, disorder-induced defects, with no dependence on the number of steps $N$, play no role in Eq.~\eqref{eq:opt_Nsteps}. From Fig.~\ref{fig:FKZnoisy_L50}, displaying the density of defects as a function of the number of steps for various noise strengths $\sigma_\mathrm{noise}$, we extract $N_{\rm opt}$, the optimal number of steps that minimizes the number of defects. In Fig.~\ref{fig:FKZoptimal_L50}, we plot $N_\mathrm{opt}$ versus $\sigma_\mathrm{noise}$ and compare it to Eq.~\eqref{eq:opt_Nsteps}. Using estimates from Fig.~\ref{fig:FKZnoise_scaling} for $a_\mathrm{ideal}$ and $a_\mathrm{noise}$, we get $(a_\mathrm{ideal}/2a_\mathrm{noise})^{2/3}=0.164(8)$, which is consistent with the coefficient obtained by fitting independently the data of Fig.~\ref{fig:FKZoptimal_L50}.

\section{Realistic noise models and experimental validation}

The noise model from Eqs.~\eqref{eq:hamiltonian_noise_pm} and~\eqref{eq:hamiltonian_noise_fm} studied in Sec.~\ref{sec:noisy_model} has the advantage of being simple, with a single parameter controlling the noise strength, and the possibility of large-scale simulations thanks to its free-fermionic nature. In what follows, we aim to compare the results of this approach with established noise models that have a better clear microscopic justification, as well as experimental results.

\subsection{Stochastic Pauli error model}

The first model is based on stochastic Pauli error models. The effect of this noise on adiabatic state preparation was first considered in Ref.~\cite{PhysRevB.106.L041109}. There, it was found that an error rate $p$ leads to a noise-induced length scale $\xi_\mathrm{noise}\sim 1/p$. This expectation translates into a density defects growing linearly with $p$. Comparing this results with the simpler noise model of Eqs.~\eqref{eq:hamiltonian_noise_pm} and~\eqref{eq:hamiltonian_noise_fm} allows one to relate the parameters controlling the noise strength in the two respective models, $p\sim\sigma_\mathrm{noise}^2$.

To obtain a realistic description of the real hardware, we consider a stochastic Pauli error model with several quantum channels $\mathcal{E}_{ij}$ applied after each two-qubit gate on qubits $(i,j)$~\cite{Nielsen2011},
\begin{equation}
    \mathcal{E}_{ij}\bigl(\rho\bigr)=\sum_{{\mu,\upsilon\in[I,x,y,z]}}p_{\mu\upsilon}\sigma_{i}^{\mu}\sigma_{j}^{\upsilon}\rho\sigma_{i}^{\mu}\sigma_{j}^{\upsilon},
    \label{eq:stochastic_pauli}
\end{equation}
with $\rho$ the density matrix describing the system, $\sigma^{I,x,y,z}$ are Pauli matrices with $\sigma^{I}\equiv I$ (identity), and $p_{\mu\upsilon}$ are the error rates fulfilling $\sum_{\mu\upsilon}p_{\mu\upsilon}=1$. Note that $p_{II}$ is the fidelity and corresponds to the probability of a perfect operation.

The parameters $p_{\mu\upsilon}$ are determined using the noise reconstruction protocol described in Refs.~\cite{erhard_characterizing_2019} and~\cite{flammia_efficient_2020}. The protocol estimates the marginal rate of each set of Pauli errors by preparing the qubit register in a basis state which is sensitive to the selected Pauli error, and then repeating the cycle of two-qubit gates $m$ times. Inserted between each cycle of two-qubit gates is a set of random Pauli gates. By repeating this measurement with different sets of interleaved Paulis, the noise is tailored to be stochastic. Finally, by varying $m$, an exponential decay rate for the Pauli error can be estimated for each channel, similar to randomized benchmarking~\cite{Knill2008}. Thus, we are able to estimate the average probability of each Pauli error occurring on a two-qubit gate during the given cycle. We perform the experiment on the superconducting quantum chip Rigetti Aspen-11. We use the hardware-native one-qubit gates $\mathtt{R_z}(\varphi)=\exp(-i\varphi\pi X/2)$ and $\mathtt{R_x}(\mathbb{Z})=\exp(-i\mathbb{Z}\pi X/2)$ as well as the two-qubit gate $\mathtt{CPHASE}(\varphi)=\textrm{diag}(1,1,1,e^{i\varphi})$ in the usual basis $\{\vert{00}\rangle,\vert{01}\rangle,\vert{10}\rangle,\vert{11}\rangle\}$ for compiling the circuit of Fig.~\ref{fig:phasediagram}(a). For the $\mathtt{CPHASE}(\varphi)$ gate, the angle of $\varphi=\pi$ is characterized. Since the phase is implemented digitally, the gate error is expected to be independent of the parametric phase.

\begin{figure}[t]
	\includegraphics[width=1\columnwidth]{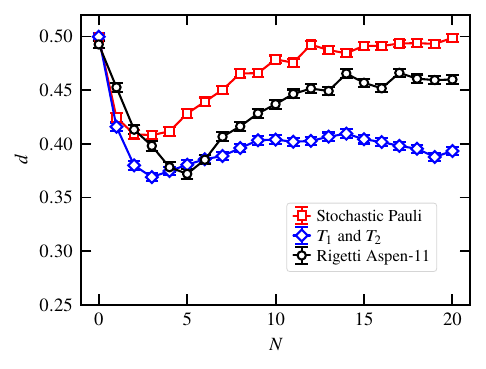}
	\caption{Density of defects versus the number of steps obtained for $L=6$ in three different ways using $2^{12}$ bitstrings for each data point: The stochastic Pauli noise model of Eq.~\eqref{eq:stochastic_pauli}, the decoherence noise model of Eq.~\eqref{eq:damping_channel}, and the experimental data from Rigetti Aspen-11. All produce phenomenologically similar curves. Note that the decoherence model data is biased downwards by the decay, especially for large number of steps. $N=0$ corresponds to a layer of Hadamard gates on individual qubits leading to $d=1/2$ from a theoretical perspective.}
	\label{fig:realistic_noise}
\end{figure}

Once estimated, the parameters can be used to simulate Eq.~\eqref{eq:stochastic_pauli}. We collect $2^{12}$ bitstrings and compute the average density of defects of Eq.~\eqref{eq:density_defects} for $L=6$ qubits versus different number of steps. We plot the results as red squares in Fig.~\ref{fig:realistic_noise} and find that this more realistic noise model shows qualitatively the same features as the simpler one previously studied. In particular, the density of defects shows a minimum for an optimal number of steps before increasing again to saturation. This result confirms that the noise model of Eqs.~\eqref{eq:hamiltonian_noise_pm} and~\eqref{eq:hamiltonian_noise_fm} is phenomenologically valid. 

\subsection{$T_1$ and $T_2$ error model}

The stochastic Pauli error model noise model of Eq.~\eqref{eq:stochastic_pauli} put the individual qubit states $\vert{0}\rangle$ and $\vert{1}\rangle$ on the same footing by acting isotropically on the Bloch sphere. However, relaxation will make qubits decay from their excited state $\vert{1}\rangle$ to their ground state $\vert{0}\rangle$, and thus favor $\vert{0}\rangle$ qubit states in the output bitstrings. As a result, the measured density of defects of Eq.~\eqref{eq:density_defects} will be lower. This effect can be modeled by an amplitude-damping channel on qubit $i$ ($T_1$), which we combine with a phase-damping channel ($T_2$)~\cite{Nielsen2011},
\begin{equation}
    \mathcal{E}_{i}\bigl(\rho\bigr)=K^1_i\rho K^{1\dag}_i + K^2_i\rho K^{2\dag}_i + K^3_i\rho K^{3\dag}_i,
    \label{eq:damping_channel}
\end{equation}
with Kraus operators $K^1_i$, $K^2_i$, and $K^3_i$,
\begin{equation}
    \begin{aligned}
        K^1_i&=\begin{pmatrix}
            1 & 0\\
            0 & \sqrt{1-\gamma_i-\bigl(1-\gamma_i\bigr)\lambda_i}
        \end{pmatrix},~~
        K^2_i=\begin{pmatrix}
            0 & \sqrt{\gamma_i}\\
            0 & 0
        \end{pmatrix},\\
        K^3_i&=\begin{pmatrix}
            0 & 0\\
            0 & \sqrt{\bigl(1-\gamma_i\bigr)\lambda_i}
        \end{pmatrix},~~\gamma_i,~\lambda_i\in[0,1],
    \end{aligned}
    \label{eq:kraus_operators}
\end{equation}
where $\gamma_i=1-\exp(-\delta t/T_{1i})$ and $\lambda_i=1-\exp(-\delta t/T_{2i})$ with $T_{1i}$ and $T_{2i}$ the relaxation and dephasing times of qubit $i$, and $\delta t$ the operation time. For the six qubits used experimentally on Rigetti Aspen-11, the average values are: $T_1\approx 30.1~\mu\mathrm{s}$ and $T_2\approx 14.5~\mu\mathrm{s}$. The gates are sorted into cycles with $\delta t\approx 32$ ns for the one-qubit gate cycle and $\delta t\approx 176$ ns for the two two-qubit gate cycles (one cycle for even and one cycle for odd bonds). We find in Fig.~\ref{fig:realistic_noise} that such $T_1$-$T_2$ error model is qualitatively similar to the others and leads to a density of defects showing a minimum for an optimal number of steps, before increasing again to saturation. However, unlike the simpler noise model and the stochastic Pauli error model, the saturation at large $N$ is obtained at $d\approx 0.5$, which is much lower than the theoretical value for a random state $d=1/2$. This behavior is expected from the asymmetry between the qubit states $\vert{0}\rangle$ and $\vert{1}\rangle$ of the noise model: The decay process described by $T_1$ drives the qubits to the $|0\rangle$ state and effectively reduces the number of defects in the chain.

\subsection{Experimental verification}

Finally, we compare the output of the noise models with the results of a real quantum computer, Aspen-11 by Rigetti. The average density of defects, Eq.~\eqref{eq:density_defects} is plotted in Fig.~\ref{fig:realistic_noise} versus the number of steps $N$, for $L=6$ qubits. We observe the non-monotonous behavior predicted by the theoretical models, with a minimal number of defects at $N=5$ steps. Interestingly, this value is larger than the one predicted by both the stochastic Pauli and $T_1$-$T_2$ realistic noise models. This effect may be attributed to correlations between the real noise occurring in the different layers, which are not captured by the theoretical models. For a long number of steps $N\gtrsim 15$, the number of defects saturates to an intermediate value $d\approx 0.45$ between the two theoretical models, respectively giving $d\approx 0.5$ and $d\approx 0.4$. These observations indicate that, while the predicted minimum is a universal effect, neither theoretical models are sufficient to achieve a quantitative description of the experiment.

\begin{figure}[t]
	\includegraphics[width=1\columnwidth]{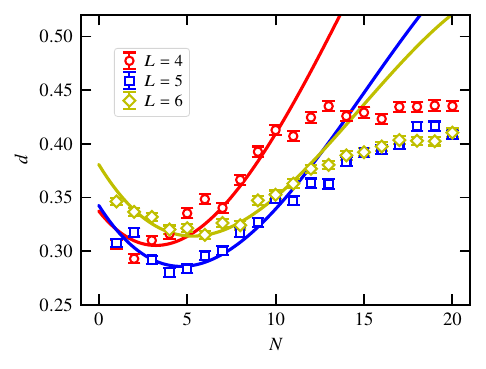}
	\caption{Experimental density of defects computed on Rigetti Aspen-M-1 for $L=4$, $L=5$, and $L=6$ from $2^{12}$ bitstrings for each data point. Lines are fit of the form $d=d_\mathrm{ideal}+\sigma^2_\mathrm{noise}N+\sigma^2_\mathrm{disorder}$ for $N\leq 11$ with $d_\mathrm{ideal}$ the density of defects in an ideal simulation. Here, $\sigma_\mathrm{noise}$ and $\sigma_\mathrm{disorder}$ are fitting parameters, see Eq.~\eqref{eq:defects_total}.}
	\label{fig:experiments}
\end{figure}

\begin{figure}[t]
    \includegraphics[width=1\columnwidth]{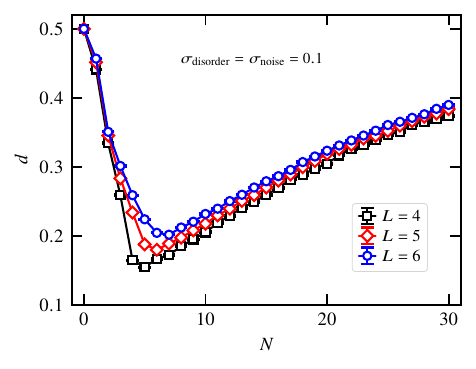}
    \includegraphics[width=1\columnwidth]{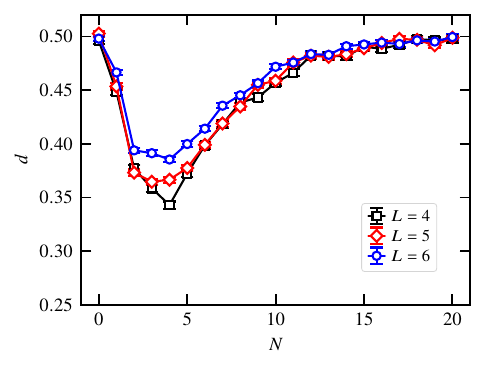}
    \caption{Density of defects $d$ as a function of the number of steps $N$ for different small system sizes $L$. Top panel: Noise model of Eqs.~\eqref{eq:hamiltonian_noise_pm} and~\eqref{eq:hamiltonian_noise_fm} with $\sigma_\mathrm{noise}=\sigma_\mathrm{disorder}=0.1$. The data points are averaged over $5000$ random realizations. Bottom panel: Stochastic Pauli noise model of Eq.~\eqref{eq:stochastic_pauli} with each data point averaged over $2^{13}$ bit strings.}
    \label{fig:noise_vs_L}
\end{figure}

To check the resilience of these results to the details of the noise source, we repeat the same experiment on a different quantum device, the Rigetti Aspen-M-1, where we consider $L=4$, $L=5$, and $L=6$ qubits, see Fig.~\ref{fig:experiments}. We find that all the system sizes follow the non-monotonous behavior predicted by the theoretical models. The optimal number of steps increases with system size from $N_\mathrm{opt}=2$ for $L=4$ to $N_\mathrm{opt}=6$ for $L=6$. By fitting the data according to Eq.~\eqref{eq:defects_total}, we can extract an effective noise strength for each system size independently. Because this expression does not capture the saturation of the density of defects at large $N$, we restrict the fitting window to $N\leq 11$ before the saturation takes place. We find the following fitting parameters: For $L=4$, $\sigma_\mathrm{noise}=0.180(6)$ and $\sigma_\mathrm{disorder}=0.27(3)$; for $L=5$, $\sigma_\mathrm{noise}=0.155(5)$ and $\sigma_\mathrm{disorder}=0.076(9)$; for $L=6$, $\sigma_\mathrm{noise}=0.142(5)$ and $\sigma_\mathrm{disorder}=0.11(1)$. This result indicates that both  $\sigma_\mathrm{disorder}$ and $\sigma_\mathrm{noise}$ decrease with the number of qubits, as reflected by an increasing $N_\mathrm{opt}$ as a function of $L$. At first, this result seems counter-intuitive because one usually expects superconducting circuits to become noisier as the system size is increased, while we observe an opposite effect. This phenomenon is also observed in the numerical simulations of Fig.~\ref{fig:noise_vs_L}. The simulations also reveal that the increase of $N_\mathrm{opt}$ with $L$ comes with an absolute larger density of defects, and from that perspective, the larger systems are not necessarily less noisy. This effect was not observed in the previous larger-scale simulations and is attributed to finite-size effects.

A similar non-monotonic behavior of the density of defects has also been recently observed in a quantum annealing experiment leveraging $2,000$ qubits of a D-Wave quantum processor, see Fig.~2(a) of Ref.~\cite{King2022} (see also Ref.~\cite{PhysRevLett.124.090502}): The density of defects initially decreases as a function of the anneal time (which plays the role of the number of steps in our discrete algorithm) and then increases back to a saturation point. The saturation value was shown to depend on the temperature of the quantum processor, which was used as an independent knob. As mentioned earlier, our noise model is equivalent to an infinite temperature bath, which drives $d$ to the saturation value of $1/2$. For low-temperature baths, the saturation value is smaller and determined by the relevant Boltzmann statistics. At very low temperatures, or equivalently at large values of $J_z$, the saturation point goes below the level of experimental detectability and the effect of the bath is rather described by a decay process analogous to the $T_1$-$T_2$ model introduced above. This experiment demonstrates the validity of our approach across different protocols.

\section{Conclusion}

\subsection{Summary}

In this work, we investigated the density of defects in the final state of a noisy, adiabatic, state-preparation circuit. On the one hand, one wants the number of layers in the circuit to be as large as possible to be in the genuine adiabatic limit and get the final state most accurately. On the other hand, inherent hardware noise will induce defects in the state preparation with each additional layer. To address this interplay in a simple scenario, we considered the evolution of a paramagnetic ground state to a ferromagnetic state in one dimension, by interpolating their respective parent Hamiltonians in $N$ steps. We found that the density of defects $d$, characterized by the density of domain walls according to Eq.~\eqref{eq:density_defects}, takes a simple form adding up two contributions $d=d_\mathrm{ideal}+d_\mathrm{noise}$: (i) A contribution from the noiseless ideal case due to the finite number of layers $N$, and (ii) a contribution from the noise of strength $\sigma$. We introduced noise in the form of a random component to the parent Hamiltonians and simulated numerically up to hundreds of qubits and thousand of steps thanks to the model mapping to free fermions.

In the noiseless case, the density of defects is controlled by the KZ mechanism with $d_\mathrm{ideal}\sim 1/\sqrt{N}$, which goes to zero in the adiabatic limit as $N\to+\infty$. We studied two versions of the noise. The first one was step- and qubit- dependent and led to a density of defects contribution proportional to $d_\mathrm{noise}\sim N\sigma_\mathrm{noise}^2$. We found that a simple random walk argument on the Bloch sphere could explain the scaling. The second version of noise was only qubit-dependent and therefore analogous to a disordered system. In that case, we observed that $d_\mathrm{disorder}\sim\sigma_\mathrm{disorder}^2$, independently of the number of steps $N$. The free-fermionic nature of the system leads to Anderson localization in presence of disorder, with a localization length going as $\xi_\mathrm{loc}\sim\sigma_\mathrm{disorder}^{-2}$, explaining the scaling of the density of defects through $d_\mathrm{disorder}\sim\xi_\mathrm{loc}^{-1}$. By obtaining a functional form for the density of defects as a function of the number of steps and the disorder strength, we derived an expression for the optimal number of steps $N_\mathrm{opt}$ minimizing the overall density of defects. We arrived to $N_\mathrm{opt}\sim\sigma_\mathrm{noise}^{-4/3}$, which we verified numerically.

We, next, considered two realistic noise models based, respectively on Pauli matrices and $T_1-T_2$ dissipative processes. These models reproduced the non-monotonous behavior of the number of defects, highlighting the universal nature of this effect. Finally, we confronted the results of the noise models with those of actual noisy quantum computers. We realized the circuit on the superconducting chips Rigetti Aspen-11 and Rigetti Aspen-M-1, and found a good agreement, validating the phenomenology of the noise model. By fitting the experimental density of defects to the functional form established in this work, we showed that one can benchmark noisy quantum processors by extracting their effective noise strength $\sigma$. This allows for an easy comparison of the performance of different hardware. 

\subsection{Outlook}

The building blocks of the quantum circuit studied in this work are the same as for the QAOA algorithm~\cite{Farhi2014,Farhi2014b,Farhi2016}: Instead of being fixed by the interpolation, the angles $J^z$ and $h^z$ are variational parameters optimized such that the final state minimizes the energy of the desired Ising Hamiltonian. Due to the similarities between QAOA and a circuit for adiabatic state preparation, we believe our results naturally extend to that case. In fact, the free fermionic nature of the circuit allows its lossless compression to a depth scaling linearly with the number of qubits~\cite{Kokcu2021,PhysRevA.105.032420,Camps2021}, showing that a QAOA-like circuit with depth $O(L)$ can represent any $N$-step adiabatic protocol. Although the numerical prefactors may be different, we expect that the predicted power-law dependence between $N_\mathrm{opt}$ and $\sigma$ should still be valid in this case.

Our results do not extend straightforwardly to higher dimensions and disordered Hamiltonians, corresponding to so-called spin glass problems~\cite{Edwards1975,RevModPhys.58.801,Castellani2005,Kawashima2013}. First, the definition of a defect based on a domain wall, as in Eq.~\eqref{eq:density_defects}, is specific to one dimension. A generalized quantity would be the excess of energy with respect to the exact ground state energy, but would require prior knowledge or estimation of the exact ground state energy. Second, the critical exponents governing the ideal noiseless case would be different depending on the dimensionality of the problem. For instance, in two dimensions, the critical exponents would be those of the Ising universality class in $(2+1)$ dimensions~\cite{PhysRevD.60.085001,PELISSETTO2002549,Kos2016,Komargodski2017}, for which a KZ mechanism has been confirmed experimentally in a cold atom setup~\cite{Ebadi2021} and numerically by neural-network-based simulations of quantum dynamics~\cite{Schmitt2021}. Finally, one would need to investigate whether the effect of noise in inducing an excess of energy $\sim N\sigma_\mathrm{noise}^2$ according to Eq.~\eqref{eq:defects_noise} remains valid beyond one dimension. We note that a qubit-dependent and step independent noise ``disorder'' would probably have a very different effect as Anderson localization is a unique property of noninteracting models (the mapping to free fermions is only valid in one dimension), and that the existence of its many-body counterpart, namely the many-body localization phenomenon~\cite{Nandkishore2015,RevModPhys.91.021001,ALET2018498}, is still actively debated beyond one dimension~\cite{Foo2022}.

Outside of quantum computing, and as discussed in the main text in relation to disorder-induced defects, we believe that interesting theoretical questions remain regarding the KZ mechanism across infinite-randomness critical points.

\begin{acknowledgments}
    This work was supported by Rigetti Computing. DA and EGDT were supported by the Israel Science Foundation, grants number 151/19 and 154/19. The experimental results presented here are based upon work supported by the Defense Advanced Research Projects Agency (DARPA) under agreement No. HR00112090058. 
\end{acknowledgments}


\begin{thebibliography}{81}
\bibitem [{kic()}]{kick}%
	\BibitemOpen
	\href@noop {} {}\bibinfo {note} {Incidentally, the model is also equivalent
		to a quantum Ising model in a time-dependent magnetic field, where the
		magnetic field is given by a periodic train of kicks.}\BibitemShut {Stop}%
\end{thebibliography}
\end{document}